RESEARCH ARTICLE

# Prediction of Metabolic Pathway Involvement in Prokaryotic UniProtKB Data by Association Rule Mining

Imane Boudellioua[1☯], Rabie Saidi[2☯]*, Robert Hoehndorf[1], Maria J. Martin[2], Victor Solovyev[3]

**1** Computational Bioscience Research Center (CBRC), Computer, Electrical and Mathematical Sciences and Engineering Division (CEMSE), King Abdullah University of Science and Technology (KAUST), Thuwal, Kingdom of Saudi Arabia, **2** European Molecular Biology Laboratory, European Bioinformatics Institute (EMBL-EBI), Cambridge, United Kingdom, **3** Softberry Inc., 116 Radio Circle, Suite 400, Mount Kisco, NY 10549, United States of America

☯ These authors contributed equally to this work.
* rsaidi@ebi.ac.uk

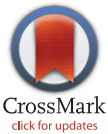

## Abstract

The widening gap between known proteins and their functions has encouraged the development of methods to automatically infer annotations. Automatic functional annotation of proteins is expected to meet the conflicting requirements of maximizing annotation coverage, while minimizing erroneous functional assignments. This trade-off imposes a great challenge in designing intelligent systems to tackle the problem of automatic protein annotation. In this work, we present a system that utilizes rule mining techniques to predict metabolic pathways in prokaryotes. The resulting knowledge represents predictive models that assign pathway involvement to UniProtKB entries. We carried out an evaluation study of our system performance using cross-validation technique. We found that it achieved very promising results in pathway identification with an $F_1$-measure of 0.982 and an AUC of 0.987. Our prediction models were then successfully applied to 6.2 million UniProtKB/TrEMBL reference proteome entries of prokaryotes. As a result, 663,724 entries were covered, where 436,510 of them lacked any previous pathway annotations.

## Introduction

One of the central research goals of systems biology is modelling various biological processes. Elucidation of chemical reactions and pathways is one of the challenging problems in this field. A biological pathway is formed by a series of chemical reactions catalyzed by enzymes within a cell. Some of the most common biological pathways are those associated with metabolism, regulation of gene expression and transmission of molecular signals. A metabolic pathway involves the step-by-step modification of an initial molecule to form another product. The resulting product can be stored by the cell, secreted, used immediately, or used to initiate another metabolic pathway. An example of a metabolic pathway is the cellular respiration

PLOS ONE

Prediction of Metabolic Pathway Involvement in Prokaryotic UniProtKB Dataequation where glucose is oxidized by oxygen to produce ATP, adenosine triphosphate [1]. Pathways play a key role in advanced studies of functional genomics. For instance, identifying pathways involved in a disease may lead to effective strategies for diagnosing, treating and preventing diseases. Moreover, by comparing the behaviour of certain pathways between a healthy person and a diseased person, researchers can discover the roots of the disorder and use the information gained from pathway analysis to develop new and better drugs [2–4]. It is increasingly clear that mapping dysregulated pathways associated with various diseases is crucial to fully understand these diseases [5]. Moreover, studying conserved pathways in model organisms may help us understand their mechanisms in other organisms.

The widening gap between the amount of known proteins and knowledge of their functions has encouraged the development of methods to automatically infer annotations. Automatic functional annotation of proteins is expected to meet the conflicting requirements of maximizing annotation coverage while minimizing erroneous functional assignments. This trade-off imposes a great challenge in designing intelligent systems to tackle the problem of automatic protein annotation. Hence, the need for automated methods is more than urgent to help increase the annotation coverage, detect inconsistencies and provide seeds for manual curation. There are several approaches proposed in the literature for such a task. A quite promising approach is to apply knowledge discovery and data mining techniques to predict some protein features based on a set of known data. Such rule-based methods provide rich automatic functional annotations and aid in performing integrity checks. For instance, Kretschmann et al [6] applied C4.5 data mining algorithm [7] to gain knowledge about the Keyword annotation from UniProtKB/Swiss-Prot [8]. Rule-base [9] is another semi-automatic annotation system run on UniProtKB/TrEMBL [8]. It uses the annotation of UniProtKB/Swiss-Prot entries that possess a set of sequence signatures to annotate UniProtKB/TrEMBL entries that contain the same signature, fundamentally with keywords and comments. Other examples of automatic annotation systems that generate annotations of several protein features integrated in UniProtKB/TrEMBL are HAMAP-Rule [10], EDIT to UniProtKB/TrEMBL [11], and PIR [12].

Literature-based manual annotation of pathways cannot scale to timely process thousands of recently sequenced genomes. Therefore, computational methods are needed for the identification and mapping of pathways. We suggested that association rule mining could be used effectively as a computational method for pathway prediction. Association rule mining is a technique originating from the analysis of data on market baskets. The objective is to locate trends by means of association relationships and correlations within a dataset. Essentially, the aim of such analysis is to discover a set of useful rules that are shared among a percentage of the dataset.

An association rule is an implication expression of the form $X \Rightarrow Y$ where X and Y are disjoint itemsets. Association rule mining was used in several applications of bioinformatics including mining gene expression data [13] and identifying related Gene Ontology terms [14]. Moreover, association rule mining was used to improve the quality of automatically generated annotations by detecting anomalies in annotation items [15]. In the context of automated protein annotation, we consider association rules in the form of many-to-one implications. If an annotation satisfies a rule with accepted quality of metric values, then we hypothesize that such a rule may reflect a biological regularity. An example of an association rule in a database of annotated proteins is: "Nuclear localization $\Rightarrow$ Origin:eukaryota", which describes that every protein which is annotated as localized in nucleus has a eukaryotic origin [15].

One of the very first pathway prediction systems was PathFinder [16] which aims to identify signaling pathways in protein-protein interaction networks. It extracts the characteristics of known signal transduction pathways and their functional annotations in the form of association rules. There are also tools that predict biodegradation pathways such as META ([17],

PLOS ONE | DOI:10.1371/journal.pone.0158896    July 8, 2016    2 / 16

CATABOL [18] and UM-PPS [19]. In addition, relative reasoning has been used for the prediction of mammalian detoxification pathways in order to limit combinatorial explosion [20]. Association rule mining was used in [21] to mine the rules linking enzymes and the domains of various genes. The Pathologic component of the Pathway Tools software [22] is the state of the art in pathway prediction. It performs predictions of metabolic pathways in sequenced and annotated genomes using MetaCyc as the reference metabolic pathway database. One of the limitations of this system is extendibility due to the fact that its logic is hardcoded. That is because Pathologic incorporates rules and heuristics developed using feedback from biologists to improve the accuracy of the predictions. Another limitation is becoming more apparent with the growth of MetaCyc in size, resulting in PathoLogic suffering from more false positive pathway predictions. In addition, the algorithm is limited to Boolean predictions with a coarse measure of prediction confidence making it difficult to filter the predictions with a probability cutoff. A comparative analysis was conducted [23] to discuss these limitations. It revealed that some machine learning approaches performed better than Pathologic in pathway prediction.

In this work, we are tackling the problem of pathway prediction in the context of metabolism. We introduce a pathway prediction system that can be used to enhance the quality of automatically generated annotations as well as annotating proteins with unknown function. The pathway prediction system utilizes data from UniProtKB/Swiss-Prot [8], which is a high quality manually annotated and non-redundant protein sequence database containing experimental results, computed features and scientific conclusions. Our pathway prediction system uses InterPro [24] signatures and organism taxonomy attributes of UniProtKB/Swiss-Prot entries to predict metabolic pathways associated with each protein entry. The association algorithm, Apriori [25], is used at the learning phase to identify significant relationships between the attributes of UniProtKB/Swiss-Prot annotations. Furthermore, we use a filtering method, SkyRule [26, 27], to select the best rules based on a combination of several interestingness metrics. This approach adopts the notions of comparability and dominance between association rules to discover the most interesting ones without favouring or excluding any measure. The selected rule will form our prediction models for pathway annotations, which can be used either directly by applying them to protein entries (using an evidence code marking it as electronically inferred), or to provide suggestions for curators to improve the annotations. We finally present an evaluation study on UniProtKB prokaryotic entries to demonstrate the performance, capability and robustness of our approach. A Java Archive (JAR) package for applying the prediction models on UniProtKB/TrEMBL prokaryotic entries is provided at http://www.ebi.ac.uk/~rsaidi/arba/software.

## Materials and Methods

### System Design

The system is designed to solve the following problem: given a set of UniProtKB/Swiss-Prot entries, generate models for pathway prediction using rule mining techniques. As with any machine learning system, the system has two major phases, the learning phase, and the application phase. The learning phase involves the training step and the testing, or validation, step on UniProtKB/Swiss-Prot input data, whereas the application phase involves applying the generated pathway models on the respective UniProtKB/TrEMBL entries. It is important to note that UniProtKB/Swiss-Prot and UniProtKB/TrEMBL sets of entries are mutually exclusive and do not overlap and therefore are selected for training and application purposes respectively. It is also worth mentioning that the system evaluation and validation is essential for the development of a reliable machine learning system. We used cross-validation as a means for system evaluation which is described later in this section.

Table 1. Current status in UniProtKB for prokaryotes.

|  | Swiss-Prot | TrEMBL |
|---|---|---|
| Total number of entries | 351,649 | 34,356,770 |
| Percentage of entries with pathway annotations | 30.44% | 5.22% |
| Percentage of entries with InterPro annotations | 98.76% | 76.17% |

As of November 2015.

doi:10.1371/journal.pone.0158896.t001

## Dataset preparation

The current status in UniProtKB for prokaryotes is summarized in Table 1, which shows that InterPro covers over 75% of prokaryotic entries in UniProtKB/TrEMBL. This high coverage will aid us in the learning process by using InterPro signatures identifiers as an attribute for the prediction models. Firstly, the system loads all prokaryotic protein entries from UniProtKB/Swiss-Prot. After that, we filter out the entries that do not contain pathway functional annotation as an attribute. Moreover, in order to maintain data quality, the system only considers entries with manual assertion evidences. An evidence is described by a code from the Evidence Codes Ontology (ECO) [28]. ECO is a controlled vocabulary of terms that describe scientific evidences in the realm of biological research. ECO can be used to document both the evidence that supports a scientific conclusion and how that conclusion was recorded by a scientist. The evidence types that are used in UniProtKB for manual assertion are described in Table 2.

The system extracts necessary information from the loaded UniProtKB/Swiss-Prot entries using metabolic pathways as targets, and InterPro signatures and organism taxonomic lineages as attributes. We ended up with a total of 96,280 entries. The attributes and target representation in UniProtKB are described as follows:

- *Target: Metabolic Pathway Comment*
  Represented as a structured hierarchy of controlled vocabulary where each process is split up into super-pathway, pathway and/or sub-pathway. When known, the step number mediated by the protein within the pathway is also indicated. On the other hand, when the metabolic pathway is not fully known, only the super-pathway and pathway labels are indicated. Moreover, a protein can participate in different pathways or in different steps of the same pathway. An example of a fully known pathway representation in UniProtKB for the protein

Table 2. Considered evidences for pathway annotation in UniProtKB/Swiss-Prot.

| Evidence ID | Evidence Label | Description |
|---|---|---|
| ECO:0000269 | Experimental evidence | Manually curated information for which there is published experimental evidence. |
| ECO:0000303 | Non-traceable author statement evidence | Manually curated information that is based on statements in scientific articles for which there is no experimental support. |
| ECO:0000305 | Curator inference evidence | Manually curated information which has been inferred by a curator based on his/her scientific knowledge or on the scientific content of an article. |
| ECO:0000250 | Sequence similarity evidence | Manually curated information which has been propagated from a related experimentally characterized protein. |
| ECO:0000255 | Sequence model evidence | Manually curated information which has been generated by the UniProtKB automatic annotation system or by various sequence analysis programs that are used during the manual curation process and which has been verified by a curator. |
| ECO:0000244 | Combinatorial evidence | Manually curated Information inferred from a combination of experimental and computational evidence. |

doi:10.1371/journal.pone.0158896.t002

Table 3. Examples of itemsets corresponding to some UniProt/Swiss-Prot entries of some prokaryotes with manual assertion evidence for pathway annotations.

| Entry ID | Corresponding Itemset |
|---|---|
| Q8TRZ4 | PATHWAY: One-carbon metabolism; methanogenesis from acetate, TAXON:Archaea, TAXON: Euryarchaeota, TAXON: Methanomicrobia, TAXON:Methanosarcinales, TAXON: Methanosarcinaceae, TAXON: Methanosarcina, IPR: IPR017896, IPR:IPR017900, IPR: IPR004460, IPR:IPR004137, IPR: IPR009051, IPR: IPR011254, IPR: IPR016099 |
| P18335 | PATHWAY: Amino-acid biosynthesis; L-arginine biosynthesis; N(2)-acetyl-L-ornithine from L-glutamate: step 4/4, PATHWAY: Amino-acid biosynthesis; L-lysine biosynthesis via DAP pathway; LL-2, 6-diaminopimelate from (S)-tetrahydrodipicolinate (succinylase route): step 2/3, TAXON: Bacteria, TAXON: Proteobacteria, TAXON: Gammaproteobacteria, TAXON: Enterobacteriales, TAXON: Enterobacteriaceae, TAXON: Escherichia, IPR:IPR017652, IPR: IPR004636, IPR:IPR005814, IPR: IPR015424, IPR:IPR015421, IPR: IPR015422 |

doi:10.1371/journal.pone.0158896.t003

*Anthranilate synthase component 1* is:*Amino-acid biosynthesis; L-tryptophan biosynthesis; L-tryptophan from chorismate: step 1/5.*

- **Attribute: InterPro Signature ID**
  The InterPro signature IDs are cross-referenced from the InterPro database, which is an integrated resource of protein families, domains and functional sites. InterPro provides a functional analysis of proteins by classifying them into families and domains. Protein signatures are combined from 11 member databases into a single searchable resource. A protein entry could be associated with one or more InterPro IDs. An example of InterPro IDs associated with the protein *Anthranilate synthase component 1* is: *IPR005801 (a domain), IPR019999 (a family), IPR006805 (a domain), IPR005256 (a family), and IPR015890 (a domain)*

- **Attribute: Taxonomic Lineage**
  The taxonomic lineage is considered as an attribute. UniProtKB Taxonomy is based on the NCBI taxonomy database and is organized in a tree structure that represents the taxonomic lineage. It contains the taxonomic hierarchical classification lineage of the source organism. It lists the nodes as they appear top-down in the taxonomic tree, with the more general grouping listed first. An example of taxonomic lineage representation for protein *Anthranilate synthase component 1* is: *Bacteria; Proteobacteria; Gammaproteobacteria; Pseudomonadales; Pseudomonadaceae; Pseudomonas.*

The extracted list of attributes and targets for each loaded entry will form an itemset. Table 3 describes some examples of the forms of itemsets that are associated with some UniProtKB/Swiss-Prot protein identifiers.

### Generation of association rules

The prepared itemsets form the input of Apriori algorithm proposed by Agarwal and Srikant [25]. Apriori, a bottom up approach, is one of the well known association rule mining techniques. Apriori discovers all significant association rules that represent trends in a large database of entries or transactions. We use the Apriori implementation developed by Borgelt [29]. This implementation uses a prefix tree to organize the support counters, and a doubly recursive procedure to process the transaction, therefore counting the support of candidate itemsets. Apriori could be configured to provide different evaluating measures for each generated association rule. Each evaluation measure tries to quantify the dependency between the antecedent and the consequent of the rule. Table 4 displays the threshold values we considered for Apriori. We use a combination of four measures to effectively minimize false positives and the number of rules generated out of pure randomness. These chosen metrics are:

Table 4. Apriori threshold values considered for the system.

| Parameter | Value |
|---|---|
| Minimum number of items per association rule | 2 |
| Minimum support of an itemset (absolute number of transactions) | 20 |
| Minimum confidence of a rule as a percentage | 100% |

doi:10.1371/journal.pone.0158896.t004

- *Support*

  According to [30], the support of an association rule $R = A$ AND $B \Rightarrow C$ (noted $supp(R)$) is the support of the set $S = A, B, C$ which is defined by the absolute or relative number of cases in which the rule is correct. In the previous example, it is the number of cases where the occurrence of item $C$ follows from the occurrences of items $A$ and $B$. However, this definition may cause some problems if multiple evaluation measures are used [29]. Hence, we will adopt the definition proposed by [29, 31, 32] which describes the support of an association rule as the absolute or relative number of cases in which it is applicable, in other words, in which its antecedent part holds. Unlike the original definition, the support in this case provides a useful statistical meaning of the support of a rule and its confidence [29].

- *Confidence*

  The confidence metric is used to measure the quality of a particular association rule. More intuitively, it measures the reliability of the inference made by a rule. Introduced in [30], the confidence of an association rule $R = X \Rightarrow Y$ (noted $conf(R)$), where $X$, and $Y$ are itemsets, is calculated as the support of the set of all items that appear in the rule, divided by the support of the antecedent set. More formally,

  $$\text{conf}(R) = \frac{supp(X \cup Y)}{supp(X)}$$

  In other words, the confidence of a rule is the number of cases in which the rule is correct relative to the number of cases in which it is applicable [29]. A high confidence ratio indicates that its associated rule has a high probability of correctness and thus makes correct predictions. It is worth mentioning that it is possible to obtain rules with high confidence out of pure chance. Such rules could be detected by demeriting whether the antecedent and the consequent parts of such rules are statistically independent. One of the measures that could assist with this is the lift value.

- *Lift Value*

  The lift value, or confidence quotient is basically the quotient of the posterior and the prior confidence of an association rule [29]. Mathematically speaking, the lift of a rule $R = X \Rightarrow Y$ is:

  $$\text{lift}(R) = \frac{conf(X \Rightarrow Y)}{conf(\emptyset \Rightarrow Y)}$$

  where $supp(\emptyset)$, is the number of transactions in the database. Lift measures how far from independence the antecedent and consequent are. A lift value equals to one, implies that the antecedent and the consequent are independent and that the support of a rule is expected, considering the supports of its components which renders such a rule not uninteresting. If the resulting lift value is greater than one, this implies that the presence of the antecedent items

Table 5. Examples of rules generated by Apriori along with their evaluation measures for UniProt/Swiss-Prot prokaryotic entries with manual assertion evidence for pathway annotations.

| Consequent | Antecedent | Support | Conf. | Lift | p-value |
|---|---|---|---|---|---|
| PATHWAY:Cofactor biosynthesis; adenosylcobalamin biosynthesis | IPR:IPR003705 | 3.24709e-04 | 1 | 90.5787 | 6.47155e-63 |
| PATHWAY:tRNA modification; archaeosine-tRNA biosynthesis | • IPR:IPR004804<br>• IPR:IPR002616<br>• TAXON:Archaea | 3.35184e-04 | 1 | 2983.44 | 2.72224e-127 |
| PATHWAY:Amino-acid biosynthesis; L-leucine biosynthesis; L-leucine from 3-methyl-2-oxobutanoate: step 2/4 | • IPR:IPR004430<br>• IPR:IPR018136<br>• IPR:IPR001030<br>• TAXON:Enterobacteriaceae<br>• TAXON:Proteobacteria<br>• TAXON:Bacteria | 8.06536e-04 | 1 | 94.6184 | 1.07237e-155 |

doi:10.1371/journal.pone.0158896.t005

raises the confidence. Likewise, if the lift value is less than one, then the presence of the antecedent items lowers the confidence.

- *p-Value*
  
  In statistics, the p-value is used to measure the statistical significance of a result. Several statistical tests have been used to calculate the p-values of association rules [33, 34]. Here, we adopt the p-value computed from G-Statistic. Under independence, the G-statistic also has a $\chi^2$-distribution. The chi-squared statistic can be used to calculate a p-value by comparing the value of the statistic to a $\chi^2$-distribution. That is, the p-value is computed as the probability that the $\chi^2$-value of an association rule can be observed by chance, assuming that the antecedent and the consequent of the rule are independent [35]. The p-value is used to infer how likely the occurrence of the rule is due to a systematic effect instead of pure random chance. If a rule has a low p-value, then this rule has a low chance of occurring if its two sides are independent. Given that this rule is observed in the data, then its two sides are unlikely to be independent, and thus, the association between them is likely to be real. On the other hand, a high p-value means that the rule has a high chance of occurring even if there is no association between its two sides. Such rules should be discarded.

Given our selected dataset and parameters, Apriori successfully generated 568,006 rules in total. Some examples of rule representation along with it quality metrics are shown in Table 5.

### Selection of association rules

Apriori generates a large number of rules especially for large databases (mining irrelevant rules, etc). The expert is unable to determine the most interesting association rules and therefore make decisions based on these rules. Hence, we need an efficient evaluation of the rules to select those that are actually relevant.

The generated list of rules will be analyzed by the SkyRule software [26, 27] to select the best rules based on their respective evaluation measures. The SkyRule operator selects the rules that are supposed to be the most interesting ones, according to several measures. SkyRule utilizes the concepts of dominance and comparability to select a family of inter-independent and statistically relevant rules. We term them representative rules. In our case, the interestingness measures considered are support, confidence, lift, and p-value that were discussed in the previous subsection. The SkyRule approach adopts the notions of comparability and dominance between association rules to discover the most interesting ones without favoring or excluding any measure among the used ones. SkyRule also allows bypassing the non-trivial issue of the threshold value specification. A rule $x = (A \Rightarrow B)$ is said to be comparable to a rule $x' =$

**Table 6. Examples of prediction models obtained in the form or aggregated rules along with their evaluation measures for UniProt/Swiss-Prot prokaryotic entries with manual assertion evidence for pathway annotations.** Each rule is accompanied by its four evaluation measures and its Euclidean distance to normalized ideal metrics.

| |
|---|
| [PREDICT] PATHWAY:Quinol/quinone metabolism; 1@4-dihydroxy-2-naphthoate biosynthesis; 1@4-dihydroxy-2-naphthoate from chorismate: step 7/7 |
| [IF] [IPR:IPR022829] 0.000332364–1.0–0.030303074366431242–1.0 → 1.3927122854520324 |
| OR [IPR:IPR029069, TAXON:Cyanobacteria] 0.000332364–1.0–0.030303074366431242–1.0 → 1.3927122854520324 |
| [END] |
| [PREDICT] PATHWAY:Purine metabolism; IMP biosynthesis via de novo pathway; N(2)-formyl-N(1)-(5-phospho-D-ribosyl)glycinamide from N(1)-(5-phospho-D-ribosyl)glycinamide (formate route): step 1/1 |
| [IF] [IPR:IPR005862] 0.00232655–1.0–0.004464281262982967–1.0 → 1.4094125301401756 |
| OR [IPR:IPR001509, IPR:IPR011761] 0.000633569–1.0–0.004464281262982967–1.0 → 1.4106114385935296 |
| OR [IPR:IPR003135, IPR:IPR013815, TAXON:Enterobacteriaceae] 0.000436228–1.0–0.004464281262982967–1.0 → 1.4107512543237724 |
| [END] |

doi:10.1371/journal.pone.0158896.t006

$(C \Rightarrow D)$ if $B = D$ AND $A \cap C \neq \emptyset$. Comparability defines a kind of semantic relationship between the rules and restricts the use of the other notion *i.e.*, dominance. Specifically, the dominance between two rules must be applied only if a semantic relationship exists between them (*i.e.*, they are comparable). The dominance relationship, which is the corner stone of the SkyRule operator can be presented as follows: a rule $r$ is said to be dominated by another rule $r'$, if for all used measures, $r'$ scores better than $r$, and therefore is more relevant.

At first, for each rule, SkyRule computes the Euclidean distance to the normalized ideal metrics (**1.0** for all four quality metrics we have). After that, SkyRule sorts the set of rules in a descending order by their associated distances. The first representative rule to be selected by SkyRule will be the rule which has metrics closest to the normalized ideal metrics. This rule has proven to be undominated by any other rule of the set of candidate rules [26, 27]. SkyRule will then discard all the rules that are comparable to this representative rule from the set of candidate rules. Essentially, SkyRule will filter out rules so that only undominated and incomparable rules are left. Out of all the rules generated by Apriori, SkyRule selected 1,347 rules as representative rules for the prediction models.

### Construction of prediction models

The rules chosen by SkyRule will be aggregated to create a model for each pathway target. For example, if we have two rules of the form $A \Rightarrow C$ and $B \Rightarrow C$, then we aggregate them to a single rule such that $A$ OR $B \Rightarrow C$. The set of aggregated rules will build the final prediction models that are described in a human readable format. This process resulted in 352 prediction models. Table 6 shows some examples of the aggregated rules presented in the form of prediction models. For each rule, the antecedent set is accompanied by its four evaluation measures and its Euclidean distance to normalized ideal metrics. The prediction models for pathway annotations can be used either directly by applying them to protein entries (using an evidence code marking it as "electronically inferred"), or to provide suggestions for curators to improve the annotations. The full list of prediction models obtained is available at http://www.ebi.ac.uk/~rsaidi/arba/prokaryotapathway/learningdetails in JSON format and can be viewed using any JSON viewer.

## System evaluation

Before explaining the evaluation process of our system, we first describe the concept of prediction used here, second the type of a prediction, and third the methodology we use to evaluate our prediction models.

**Prediction and prediction types.** In our case, a prediction is defined as an association between a protein and a pathway annotation *e.g.*, the couple ("Q8TRZ4", One-carbon metabolism; methanogenesis from acetate) means that the pathway, "One-carbon metabolism; methanogenesis from acetate" has been predicted for the protein of ID "Q8TRZ4".

This association is created when a prediction model, generated by our system, is applied on a protein. When the protein satisfies the model conditions, the generated prediction is called a positive prediction or simply a prediction. Otherwise, it is called a negative prediction *i.e.*, according to the system, the protein is not involved in the pathway under consideration. Furthermore, if a prediction is generated for a protein with known pathway annotations, then this prediction can be qualified as true if it is correct or false otherwise. So, a prediction can be a true positive (TP), true negative (TN), false positive (FP) or false negative (FN). For example, if a protein has an annotation with two pathways and the system predicted only one of the two pathways, then we will count one true positive and one false negative. A more general example is: if we have $x$ number of pathways in the reference set of pathways, and $y$ proteins where each protein is annotated with a unique pathway from the set of $x$, assuming we predicted them all correctly, then we will have $y$ TP and $y(x-1)$ TN.

**Evaluation process.** For this experiment, we used the set of UniProtKB/Swiss-Prot prokaryotic entries containing pathway annotations with manual assertion evidences (96,280 entries in total as of November 2015). We define as a pathway reference set the set of pathways present in at least 20 protein entries.

In order to evaluate the robustness of our system, we used a five-fold cross-validation procedure with two runs (or rounds). In each round, data is split into five complementary folds and five iterations of learning are performed. At each iteration, both rule mining and model building are performed using four folds independently of the validation set (remaining fold). The five obtained validation results are then averaged over the number of folds to summarize the performance for one run of the evaluation process. We perform a second round of five-fold-cross-validation with a different splitting in order to avoid any bias that would result from the splitting process. The global validation result is then obtained by averaging the results over the two runs. Table 7 presents the global confusion matrix obtained from the performance analysis.

We consider four different evaluation metrics, precision, recall, $F_1$-measure and area under the ROC curve (AUC), defined as:

- $Precision = \frac{TP}{(TP+FP)}$

- $Recall = \frac{TP}{(TP+FN)}$

**Table 7. Performance evaluation of our system as illustrated by a confusion matrix.** Results are averaged over two-run five-fold-cross-validation along with the corresponding deviation values (±*d*) from the observed values of the two runs.

|  | Positive prediction | Negative prediction |
|---|---|---|
| Actually positive | TP = 109,136 ± 47 | FN = 3824 ± 53 |
| Actually negative | FP = 22 ± 2 | TN = 33,180,542 ± 1,015 |

doi:10.1371/journal.pone.0158896.t007

Table 8. Evaluation metrics of cross-validation experiment over UniProtKB/Swiss-Prot prokaryotic entries with pathway annotations of manual assertion evidence.

| Metric | Value |
| --- | --- |
| Precision | 0.999 |
| Recall | 0.966 |
| $F_1$-measure | 0.982 |
| AUC | 0.987 |

doi:10.1371/journal.pone.0158896.t008

- $F_1\text{-measure} = 2 \cdot \frac{Precision \cdot Recall}{Precision + Recall}$

- *AUC* = Area under the ROC curve. A receiver operating characteristic (ROC) curve is a plot of the true positive rate as a function of the false positive rate and is often used to illustrate the performance of a binary score-based classifier. The area under the ROC curve is a summary measure that essentially averages diagnostic accuracy across the spectrum of threshold values.

Table 8 presents the global evaluation metrics calculated over all target pathways. The table shows that our system achieved promising results in pathway identification with an $F_1$-measure of 0.982, a precision of 0.999, a recall of 0.966 and an AUC of 0.987.

## Annotation of UniProtKB/TrEMBL entries

In order to capture the performance of our system, we considered the *reference proteome* set of prokaryotic entries of UniProtKB/TrEMBL for the purpose of annotation using our prediction models. Reference proteomes are a subset of proteomes that have been selected either manually or algorithmically according to a number of criteria, to provide a broad coverage of the tree of life and a representative cross-section of the taxonomic diversity found within UniProtKB. It also covers the proteomes of well-studied model organisms and other species of interest for biomedical research [8]. These reference proteomes are tagged with the keyword "Reference proteome". As of November 2015, the reference proteome set of UniProtKB/TrEMBL entries of prokaryotes represents a fraction of around 18% over all prokaryotic UniProtKB/TrEMBL entries available in UniProtKB. In details, there are 6,193,540 prokaryotic reference proteome entries in UniProtKB/TrEMBL out of 34,356,770 total prokaryotic UniProtKB/TrEMBL entries. Table 9 summarizes some statistics about pathway annotations and InterPro annotations in our UniProtKB/TrEMBL reference proteome set. The coverage of our automatic annotations over the set specified is illustrated in the next section. Moreover, the link http://www.ebi.ac.uk/~rsaidi/arba/prokaryotapathway/organisms/comparison contains predictions applied on some popular prokaryotic organisms present in UniProtKB/TrEMBL along with a

Table 9. Current status in UniProtKB/TrEMBL for prokaryotic reference proteome set.

|  | TrEMBL |
| --- | --- |
| Total number of entries | 6,193,540 |
| Percentage of entries with pathway annotations | 3.67% |
| Percentage of entries with InterPro annotations | 80.68% |

As of November 2015.

doi:10.1371/journal.pone.0158896.t009

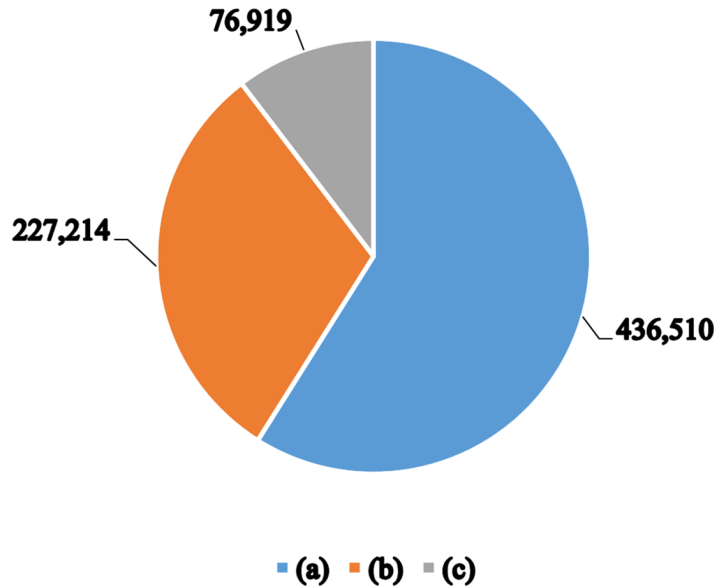

**Fig 1. Annotation coverage for UniProtKB/TrEMBL reference proteome prokaryotic entries. (a)** represents entries we could cover which lack pathway annotation, **(b)** represents entries we could cover which already have pathway annotation, and **(c)** represents entries we could not cover which already have pathway annotation.

doi:10.1371/journal.pone.0158896.g001

graphical report, illustrating our system's predictions compared to those made by other systems present in UniProtKB/TrEMBL.

## Runtime analysis

The system achieved a very high coverage of prokaryotic pathway prediction. That is, considering the parameters defined earlier in Table 4, out of 394 different pathways that are present in the learning dataset, 355 were validated and used to build pathway models for annotating UniProtKB/TrEMBL. The system took 77 minutes to generate the pathway models of all considered prokaryotic UniProtKB/Swiss-Prot entries with manual assertion evidences. The system was run on a 64-bit machine that has an Intel Core 3.00 GHz processor and a 16GB RAM.

## Results and Discussion

### Distribution of annotation coverage

Here, we provide a comparison of the annotation coverage of our system over UniProtKB/TrEMBL with reference to all other automatic annotation systems run on UniProtKB/TrEMBL such as Rule-base [9] and HAMAP-Rule [10]. Fig 1 illustrates some statistics about the UniProtKB/TrEMBL entries annotated by our system as follows. Out of 6,193,540 prokaryotic reference proteome entries in UniProtKB/TrEMBL, 663,724 were annotated using the prediction models built by our system. Interestingly, a considerably large set of 436,510 entries lacked any previous pathway annotations and is now annotated by our system. A total of 227,214 entries of those covered constitute the entries that had previous annotations by other systems, in addition to the annotation proposed by our system. It is worth mentioning that there are only 76,919 entries that had been annotated by other systems and were not annotated by our

Table 10. Overview of HAMAP-Rule, Rule-Base, and SAAS.

| System | Methodology | Evaluation methodology |
|---|---|---|
| HAMAP-Rule | Semi-automated/manual: rules are created by bio-curators and applied automatically | Bio-curator expertise: |
| Rule-Base | Semi-automated/manual: rules are created by bio-curators, statistically validated and applied automatically. | Bio-curator expertise + Each rule must have a confidence of more than 95%. |
| SAAS | Automated: rules are created by a C4.5 decision tree algorithm and applied automatically. | Each rule must have a confidence of more than 95%. |

doi:10.1371/journal.pone.0158896.t010

prediction models. Table 10 summarises how each of the three systems work. Although our system has stricter requirements to identify rules (i.e., rules must have a confidence of 100% and must pass the cross-validation process), it nevertheless can annotate more entries since our rules are directly derived from the data, not manually created, and derived from a larger dataset compared to SAAS.

## Comparison of annotation coverage

In Fig 2, we compare the annotation coverage of our system to three other main automatic annotation systems present in UniProtKB/TrEMBL, namely SAAS [7], HAMAP-Rule, and Rule-base. Our system significantly surpasses the other three systems in terms of the number of entries covered. It annotated 663,724 entries where the next best system was HAMAP-Rule with a coverage of only 229,402 entries. Rule-base annotated the least number of entries of only 93,613 entries.

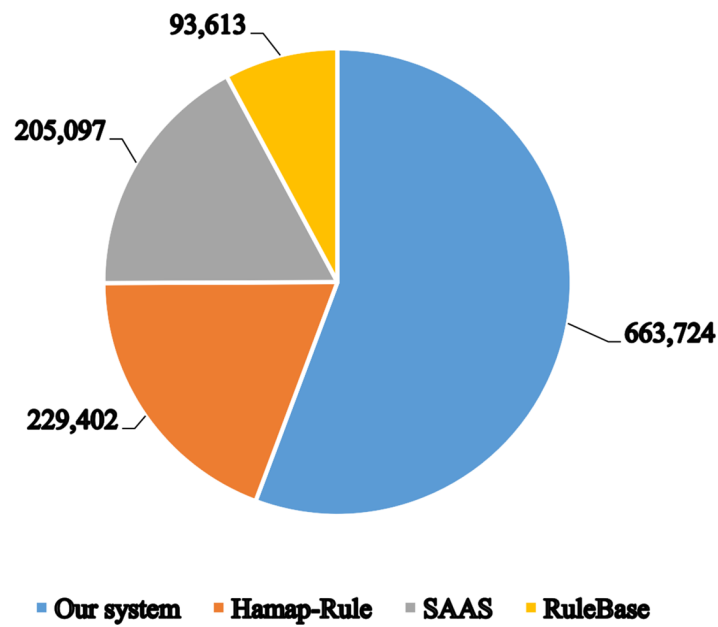

Fig 2. Comparison of annotation coverage of UniProtKB/TrEMBL reference proteome prokaryotic entries with three main automatic annotation systems present in UniProtKB/TrEMBL which are SAAS, HAMAP-Rule, and Rule-base.

doi:10.1371/journal.pone.0158896.g002

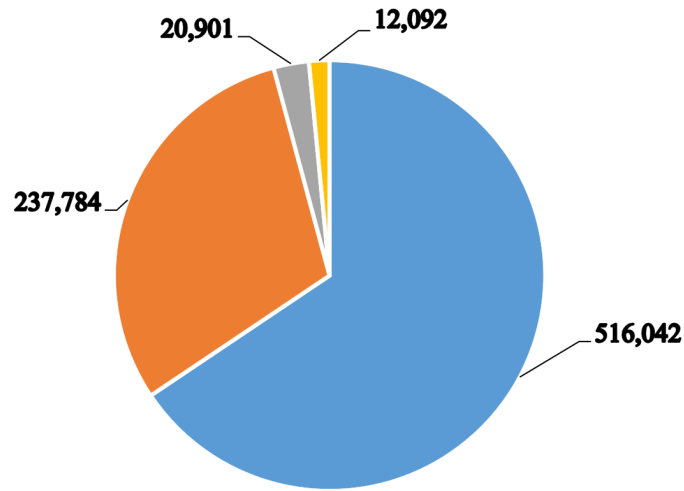

**Fig 3. Comparison of predictions applied on UniProtKB/TrEMBL reference proteome prokaryotic entries relative to three main automatic annotation systems present in UniProtKB/TrEMBL which are HAMAP-Rule, SAAS and Rule-base.**

doi:10.1371/journal.pone.0158896.g003

## Comparison of total number of prediction

In Fig 3, we take a deeper look into the various predictions made by our system in comparison to those made by Rule-base, SAAS, and HAMAP-Rule. Note that an entry in UniProtKB/TrEMBL could gain multiple predictions and hence obtain multiple pathway annotations accordingly. Here, we were able to make a total of 786,819 predictions by our system where the majority of these predictions, 516,042, were made to entries that have no previous pathway annotation. Moreover, 237,784 predictions were found to be identical matches to the annotations proposed by other systems. We also found 20,901 of our annotations similar to those proposed by other systems, either being more specific or more general in their pathway hierarchical representation. Finally, there were 12,092 predictions distinct from those already assigned by other systems.

In order to better quantify the proportion of identical or similar predictions shared between our system and the other three main automatic annotation systems, Rule-base, HAMAP-Rule and SAAS, we compare the predictions that correspond to entries annotated by our system and the three other systems. Fig 4 compares the distribution of annotations produced by our system and those provided by Rule-base, HAMAP-Rule and SAAS systems. For instance, there were 35,849 annotations in Rule-base that are identical to those predicted by our system, while there were 209,860 and 160,895 predictions identical to those made by HAMAP-Rule and SAAS respectively. On the other hand, we found 4,816 Rule-base, 19,498 HAMAP-Rule and 10,294 SAAS annotations that were similar to those annotated by our system. The similarity occurs due to the hierarchical property of pathway annotations that renders some annotations to be either more general or more specific. Moreover, we observed 4,750 Rule-base annotations, 1,201 HAMAP-Rule annotations, and 3,370 SAAS annotations that were completely different to those annotations provided by our system. These results indicate that for those entries

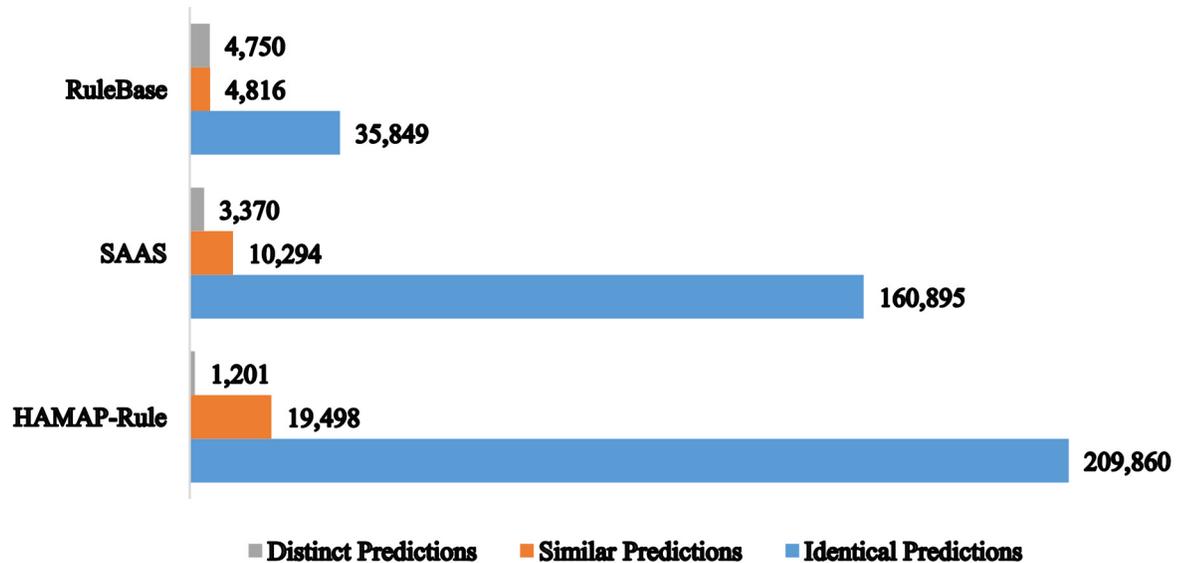

**Fig 4. Comparison of predictions corresponding to UniProtKB/TrEMBL reference proteome prokaryotic entries touched by our system, Hamap-Rule, SAAS, and Rule-base.**

doi:10.1371/journal.pone.0158896.g004

annotated by both our system and the other two systems, the majority of predictions were identical and similar. This provides an insight about the behaviour of our system as an automatic annotation tool. This shared similarity supports the validity of our prediction models and their relevance on UniProtKB/TrEMBL entries.

## Acknowledgments

The first author conducted this work as part of a research internship at the European Bioinformatics Institute, UniProt team. The authors would like to thank UniProt Consortium for their valuable support and feedback on the development of this work.

## Author Contributions

Conceived and designed the experiments: RS IB MJM VS. Performed the experiments: IB RS. Analyzed the data: IB RS RH. Contributed reagents/materials/analysis tools: RS IB MJM. Wrote the paper: IB RS. Proofread the manuscript: RS IB VS MJM RH. Developed the tool: RS IB.

## References

1. Campbell NA, Reece JB. Biology. No. v. 1 in Addison-Wesley world student series. Benjamin Cummings; 2002. Available from: http://books.google.com.sa/books?id=9pqXQgAACAAJ
2. Chen X, Xu J, Huang B, Li J, Wu X, Ma L, et al. A sub-pathway-based approach for identifying drug response principal network. Bioinformatics. 2011; 27(5):649–654. Available from: http://bioinformatics.oxfordjournals.org/content/27/5/649.abstract doi: 10.1093/bioinformatics/btq714 PMID: 21186246
3. Chen Y, Hu Y, Zhou T, Zhou KK, Mott R, Wu M, et al. Activation of the Wnt Pathway Plays a Pathogenic Role in Diabetic Retinopathy in Humans and Animal Models. The American Journal of Pathology. 2009; 175(6):2676–2685. Available from: http://www.sciencedirect.com/science/article/pii/S0002944010607754 doi: 10.2353/ajpath.2009.080945 PMID: 19893025

4. Silberberg Y, Gottlieb A, Kupiec M, Ruppin E, Sharan R. Large-scale elucidation of drug response pathways in humans. Journal of computational biology: a journal of computational molecular cell biology. 2012 Feb; 19(2):163–174. Available from: http://dx.doi.org/10.1089/cmb.2011.0264

5. Parkes M, Cortes A, van Heel DA, Brown MA. Genetic insights into common pathways and complex relationships among immune-mediated diseases. Nat Rev Genet. 2013 Sep; 14(9):661–673. Analysis. Available from: http://dx.doi.org/10.1038/nrg3502 PMID: 23917628

6. Kretschmann E, Fleischmann W, Apweiler R. Automatic rule generation for protein annotation with the C4.5 data mining algorithm applied on SWISS-PROT. Bioinformatics. 2001; 17(10):920–926. Available from: http://bioinformatics.oxfordjournals.org/content/17/10/920.abstract doi: 10.1093/bioinformatics/17.10.920 PMID: 11673236

7. Quinlan JR. C4.5: Programs for Machine Learning. San Francisco, CA, USA: Morgan Kaufmann Publishers Inc.; 1993.

8. The UniProt Consortium. UniProt: a hub for protein information. Nucleic Acids Research. 2015; 43(D1):D204–D212. Available from: http://nar.oxfordjournals.org/content/43/D1/D204.abstract doi: 10.1093/nar/gku989 PMID: 25348405

9. Biswas M, O Rourke JF, Camon E, Fraser G, Kanapin A, Karavidopoulou Y, et al. Applications of InterPro in protein annotation and genome analysis. Briefings in Bioinformatics. 2002; 3(3):285–295. doi: 10.1093/bib/3.3.285 PMID: 12230037

10. Pedruzzi I, Rivoire C, Auchincloss AH, Coudert E, Keller G, de Castro E, et al. HAMAP in 2013, new developments in the protein family classification and annotation system. Nucleic Acids Research. 2013; 41(D1):D584–D589. doi: 10.1093/nar/gks1157 PMID: 23193261

11. Muller S, Leser U, Fleischmann W, Apweiler R. EDITtoTrEMBL: a distributed approach to high-quality automated protein sequence annotation. Bioinformatics. 1999; 15(3):219–227. Available from: http://bioinformatics.oxfordjournals.org/content/15/3/219.abstract doi: 10.1093/bioinformatics/15.3.219

12. Wu CH, Huang H, Arminski L, Castro-Alvear J, Chen Y, Hu ZZ, et al. The Protein Information Resource: an integrated public resource of functional annotation of proteins. Nucleic Acids Research. 2002; 30(1):35–37. Available from: http://nar.oxfordjournals.org/content/30/1/35.abstract doi: 10.1093/nar/30.1.35 PMID: 11752247

13. Creighton C, Hanash S. Mining gene expression databases for association rules. Bioinformatics. 2003; 19(1):79–86. Available from: http://bioinformatics.oxfordjournals.org/content/19/1/79.abstract doi: 10.1093/bioinformatics/19.1.79 PMID: 12499296

14. Bodenreider O, Aubry M, Burgun A. Non-Lexical Approaches to Identifying Associative Relations in the Gene Ontology. In: Altman RB, Jung TA, Klein TE, Dunker AK, Hunter L, editors. Pacific Symposium on Biocomputing. World Scientific; 2005. p. 104–115.

15. Artamonova II, Frishman G, Gelfand MS, Frishman D. Mining sequence annotation databanks for association patterns. Bioinformatics. 2005; 21(Suppl 3):iii49–iii57. Available from: http://bioinformatics.oxfordjournals.org/content/21/Suppl_3/iii49.abstract doi: 10.1093/bioinformatics/bti1206 PMID: 16306393

16. Bebek G, Yang J. PathFinder: mining signal transduction pathway segments from protein-protein interaction networks. BMC Bioinformatics. 2007; 8(1):335. Available from: http://www.biomedcentral.com/1471-2105/8/335 doi: 10.1186/1471-2105-8-335 PMID: 17854489

17. Klopman G, Tu M, Talafous J. META. 3. A Genetic Algorithm for Metabolic Transform Priorities Optimization. Journal of Chemical Information and Computer Sciences. 1997; 37(2):329–334. Available from: http://dx.doi.org/10.1021/ci9601123 PMID: 9157102

18. Jaworska J, Dimitrov S, Nikolova N, Mekenyan O. Probabilistic assessment of biodegradability based on metabolic pathways: CATABOL System. SAR and QSAR in Environmental Research. 2002; 13(2):307–323. Available from: http://dx.doi.org/10.1080/10629360290002794 PMID: 12071658

19. Hou B, Ellis LM, Wackett L. Encoding microbial metabolic logic: predicting biodegradation. Journal of Industrial Microbiology and Biotechnology. 2004; 31(6):261–272. Available from: http://dx.doi.org/10.1007/s10295-004-0144-7 PMID: 15248088

20. Button WG, Judson PN, Long A, Vessey JD. Using Absolute and Relative Reasoning in the Prediction of the Potential Metabolism of Xenobiotics. Journal of Chemical Information and Computer Sciences. 2003; 43(5):1371–1377. Available from: http://dx.doi.org/10.1021/ci0202739 PMID: 14502469

21. Chiu SH, Chen CC, Yuan GF, Lin TH. Association algorithm to mine the rules that govern enzyme definition and to classify protein sequences. BMC Bioinformatics. 2006; 7(1):304. Available from: http://www.biomedcentral.com/1471-2105/7/304 doi: 10.1186/1471-2105-7-304 PMID: 16776838

22. Karp P, Latendresse M, Caspi R. The Pathway Tools Pathway Prediction Algorithm. Standards in Genomic Sciences. 2011; 5(3). Available from: http://standardsingenomics.org/index.php/sigen/article/view/sigs.1794338 doi: 10.4056/sigs.1794338 PMID: 22675592

23. Dale J, Popescu L, Karp P. Machine learning methods for metabolic pathway prediction. BMC Bioinformatics. 2010; 11(1):15. Available from: http://www.biomedcentral.com/1471-2105/11/15 doi: 10.1186/1471-2105-11-15 PMID: 20064214

24. The InterPro Consortium, Mulder NJ, Apweiler R, Attwood TK, Bairoch A, Bateman A, et al. InterPro: An integrated documentation resource for protein families, domains and functional sites. Briefings in Bioinformatics. 2002; 3(3):225–235. Available from: http://bib.oxfordjournals.org/content/3/3/225.abstract doi: 10.1093/bib/3.3.225 PMID: 12230031

25. Agrawal R, Srikant R. Fast Algorithms for Mining Association Rules in Large Databases. In: Bocca JB, Jarke M, Zaniolo C, editors. VLDB 94, Proceedings of 20th International Conference on Very Large Data Bases, September 12-15, 1994, Santiago de Chile, Chile. Morgan Kaufmann; 1994. p. 487–499.

26. Bouker S, Saidi R, Yahia SB, Nguifo EM. Ranking and Selecting Association Rules Based on Dominance Relationship. In: IEEE 24th International Conference on Tools with Artificial Intelligence, ICTAI 2012, Athens, Greece, November 7-9, 2012; 2012. p. 658–665. Available from: http://dx.doi.org/10.1109/ICTAI.2012.94

27. Bouker S, Saidi R, Yahia SB, Nguifo EM. Mining Undominated Association Rules Through Interestingness Measures. International Journal on Artificial Intelligence Tools. 2014; 23(4). Available from: http://dx.doi.org/10.1142/S0218213014600112

28. Chibucos MC, Mungall CJ, Balakrishnan R, Christie KR, Huntley RP, White O, et al. Standardized description of scientific evidence using the Evidence Ontology (ECO). Database. 2014; 2014. Available from: http://database.oxfordjournals.org/content/2014/bau075.abstract doi: 10.1093/database/bau075 PMID: 25052702

29. Borgelt C, Kruse R. Induction of Association Rules: Apriori Implementation. In: Proc. of the 15th Conference on Computational Statistics (COMPSTAT). Physica Verlag; 2002. p. 395–400.

30. Agrawal R, Srikant R. Fast Algorithms for Mining Association Rules in Large Databases. In: Proceedings of the 20th International Conference on Very Large Data Bases. VLDB 94. San Francisco CA, USA: Morgan Kaufmann Publishers Inc.; 1994. p. 487–499. Available from: http://dl.acm.org/citation.cfm?id=645920.672836

31. Borgelt C. Efficient Implementations of Apriori and Eclat. In: Proc. 1st IEEE ICDM Workshop on Frequent Item Set Mining Implementations (FIMI 2003, Melbourne, FL). CEUR Workshop Proceedings 90; 2003. p. 90.

32. Borgelt C. Recursion Pruning for the Apriori Algorithm. In: Jr RJB, Goethals B, Zaki MJ, editors. FIMI. vol. 126 of CEUR Workshop Proceedings. CEUR-WS.org; 2004. Available from: http://dblp.uni-trier.de/db/conf/fimi/fimi2004.html#Borgelt04

33. Brin S, Motwani R, Silverstein C. Beyond Market Baskets: Generalizing Association Rules to Correlations. In: Proceedings of the 1997 ACM SIGMOD International Conference on Management of Data. SIGMOD 97. New York NY, USA: ACM; 1997. p. 265–276. Available from: http://doi.acm.org/10.1145/253260.253327

34. Kirsch A, Mitzenmacher M, Pietracaprina A, Pucci G, Upfal E, Vandin F. An Efficient Rigorous Approach for Identifying Statistically Significant Frequent Itemsets. In: Proceedings of the Twenty-eighth ACM SIGMOD-SIGACT-SIGART Symposium on Principles of Database Systems. PODS 09. New York NY, USA: ACM; 2009. p. 117–126. Available from: http://doi.acm.org/10.1145/1559795.1559814

35. National Human Genome Research Institute NIoH. Biological Pathways;. http://www.genome.gov/27530687